\documentstyle[aps,preprint]{revtex} \begin{document}
\newcommand{\kbar}{\shortstack{\hspace*{0.02em}\_\\*[-1.0ex]\it k}}
\newcommand{\lvec}{\partial^{^{\!\!\!\!\leftarrow}}}
\newcommand{\rvec}{\partial^{^{\!\!\!\!\rightarrow}}} \draft 

\title{Sensitivity to measurement perturbation of single atom dynamics  in
cavity QED} 
\author{X.~M.~Liu , M.~Hug , and G.~J.~Milburn} 
\address{The Center for Laser Sciences, Department of Physics,} 
\address{The University of Queensland, St. Lucia, Brisbane, Qld 4072,Australia} 
\date{May 3, 2000}
\maketitle  
\begin{abstract} 

We consider continuous observation of the
nonlinear dynamics of single atom trapped in an optical cavity  by a standing wave
with intensity modulation. The motion of the atom changes the phase of the
field which is then monitored by homodyne detection of the output field. We
show that the conditional Hilbert space dynamics of this system, subject to
measurement induced perturbations, depends strongly on whether the
corresponding classical dynamics is regular or chaotic. If the classical
dynamics is chaotic the distribution of conditional Hilbert space vectors
corresponding to different observation records tends to be orthogonal. This is
a characteristic feature of hypersensitivity to perturbation for quantum
chaotic systems.  
\end{abstract} 
\narrowtext 

\section{Introduction}
The study of quantum nonlinear dynamics, especially systems which classically
exhibit Hamiltonian chaos, has recently begun to focus on the response of such
systems to external sources of noise and decoherence\cite{Zurek94,Schack94}.
This direction was prompted by the observation that nonintegrable classical
systems, when quantized will exhibit dynamics that departs from that expected
classically on a very short time scale\cite{Fishman82,Shepelyansky86,Chirikov89},
so short that even macroscopic
systems should show observable quantum features in their motion\cite{Zurek}. 
There have now been numerous experimental observations of the short time
deviations between quantum and classical dynamics of nonlinear
systems\cite{Galvez88,Bayfield89,Arndt91,Moore95}. The nonlinear dynamics of
cold atoms in optical dipole potentials has proved to be a particularly fertile
field for quantum nonlinear dynamics. Recently the effects of decoherence in
quantum chaotic dynamics was studied using cold atoms
\cite{Klappauf1,Klappauf2,Ammann98}.  However in all experimental observations
so far, the results were obtained from an ensemble of systems, not from
repeated observations on a single quantum system. Recent progress in single
atom dynamics in small optical cavities\cite{Ye} indicate that it will soon be
possible to study the quantum nonlinear dynamics of a single quantum system
subject to repeated measurements\cite{doherty99} and it is towards describing
such systems that this paper is directed.  

It is in the context of such single
system dynamics that the information approach of Schack and Caves\cite{Schack}
based on hypersensitivity to perturbation  becomes significant. In that
approach the response of classical and quantum nonlinear systems to external
perturbations is considered. In particular they show that for a chaotic system
it requires a huge  amount  algorithmic information to track the classical
(phase space trajectory) or quantum (Hilbert space vector) of a single chaotic
system when it is subjected to small external perturbations. It is better in
such cases to average over the perturbation and pay a much smaller cost in von
Neumann entropy. In the quantum case the signature of this hypersensitivity to
perturbation has been shown to be the distribution of Hilbert space vectors
resulting from dynamical sequences with different perturbation histories. While
this approach seems to offer considerable insight into quantum and classical
chaos, it is far from clear what  it means for an experiment where the dominant
source of perturbation is likely to be the measurement back action associated
with the attempt to continuously monitor the dynamics.  

In reference
\cite{Breslin} an attempt was made to study hypersensitivity to perturbation
arising from quantum measurements made on a single quantum system: a nonlinear
kicked top. In that study the measurements were not continuous in time but
rather a sequence of discrete readouts applied at the same time as the kicks.
The results confirmed in general terms the observation of Schack and Caves for
measurement induced perturbation. Specifically it was shown that  if a system
was initially localized on a chaotic region of phase space, the Hilbert space
vectors resulting from different measurement histories tended to become
orthogonal, while for initial regular states the Hilbert space vectors for
different histories tended to remain closer together.  In this paper we extend
that study to the case of a continuously monitored single quantum nonlinear
system: a single atom trapped by an intracavity optical dipole field.  The
motion of the atom changes the  phase of the cavity field which may be
monitored using phase sensitive detection of the light leaving the cavity.
Chaos is introduced by externally modulating the intensity of the light inside
the cavity.  We use the now established techniques of quantum
trajectories\cite{carmichael,molmer} to study the distribution of Hilbert space
vectors for different measurement histories. Previous studies that use quantum
trajectories to describe the dynamics of open quantum nonlinear systems include
the work of Brun et al.  \cite{Brun97}.   

We use the nonlinear stochastic
Schr\"{o}dinger equation to parallel the discussion in reference \cite{Breslin}
based on how information is extracted.  The nonlinear Schr\"{o}dinger equation 
describes a true measurement in which actual information about the quantum
state of the monitored system is extracted from the external field. Our results
confirm that a chaotic system,  subject to different continuous observation
histories,  will produce a distribution of states that tend to be orthogonal.
This means that a very tiny error in recording the measurement history will
suggest a final state that is very likely orthogonal to the actual final state.
In this way the intuitive idea that chaos constrains predictability is carried
over to continuously observed single nonlinear quantum systems.  
 
\section{theoretical model of Homodyne measurements on  single atom dynamics
in  cavity QED  }  

Recently, experiments in cavity QED have achieved the
exceptional circumstance of strong coupling, for which single quanta can impact
the atom-cavity system. The trapping of  a single atom in a high-finesse cavity
has been realized\cite{Ye}. In these systems we must treat quantum mechanically
both the optical and electronic degrees of freedom as well as the
center-of-mass motion of the atom. In our model the atom is in the optical
dipole potential of a cavity standing wave which is blue detuned from an atomic
resonance so that there is a net conservative force acting on the atom in the
direction of decreasing intensity. This interaction does not change the
intensity of the optical field but it does change the phase by an amount that
depends on the atomic position. As the atom moves in the cavity it changes the
phase of the field and if this phase change can be monitored we can effectively
monitor the atomic position. This can be accomplished by a homodyne measurement
of the field leaving the optical cavity. Mabuchi et al\cite{mabuchi} have
already demonstrated this kind of measurement at the level of a single atom.  
A similar model for an atom trapped in a harmonic optical potential was
recently discussed by Doherty et al\cite{doherty99}  

The basic theoretical
description can be given as master equation for a two-level atom coupling to a
single electromagnetic mode via the Jaynes-Cummings interaction Hamiltonian,
including the quantization of the atomic center-of-mass\cite{Herkommer,Quadt}. 
The Hamiltonian in Schr\"{o}dinger picture can be written as  

\begin{equation}
\hat{H}=\frac{\hat{p}^2}{2M}+\hbar\omega_{A}\sigma^{+}\sigma^{-} + \hbar
\omega_{c} a^{+} a  + \hbar E_{0} (a e^{-i\omega_{L} t}+ a^{+} e^{
i\omega_{L}t})+  \hbar g \sin(k_{L}\hat{x})(a\sigma^{+}+a^{+} \sigma^{-}),
\end{equation} 
where $p$ is the momentum of atom, $M$ is its mass, 
$\omega_{A}$, $\omega_{c}$, and  $\omega_{L}$ are the the two-level resonance
frequency, the cavity frequency, and the frequency of the driving laser field,
respectively. The term $E_{0}$ is a constant  proportional to the  amplitude of
the driving field, $g$ is the coupling constant of the interaction  between 
driving field and atom, $\sigma^{+}$ and $\sigma^{-}$ are the   raising and
lowering operators for the two-level atom, and   $a^{+}$ and $a$ are the
creation and annihilation operators for the   cavity field.   We assume that
the detuning $\Delta$ is positive and  $\Delta=\omega_{A}-\omega_{L} \gg  g,
\Gamma$ and $\omega_{c}=\omega_{L}$,  where $\Gamma$ is the atomic dipole decay
rate. In the interaction picture  the Hamiltonian can be simplified as   
\begin{equation} 
\hat{H}^{'}=\hat{H}_{eff}+\hbar E_{0}(a+a^{+}) 
\end{equation} 
where 
\begin{equation}
\hat{H}_{eff}=\frac{\hat{p}^2}{2M}-\frac{\hbar{g^2}}{2\Delta}a^{+}a\cos(2k_{L}\hat{x}),
\end{equation} 
is the effective Hamiltonian\cite{Herkommer}. Note that the
effective interaction does not include the driving  laser field.  

We denote
$\Lambda$ the density operator for the joint state of the atom and the cavity.
Then the master equation for $\Lambda$ is \cite{Wiseman,Corney},

\begin{equation}
\frac{d\Lambda}{dt}=\frac{1}{i\hbar}[\hat{H}_{eff},\Lambda]-iE_{0}[a+a^{+},
\Lambda]+ \frac{\kappa}{2}(2a\Lambda a^{+}-a^{+}a\Lambda-\Lambda a^{+}a),
\end{equation}
where $\kappa$ is the cavity decay rate. Note that if the cavity
is driven by a strong coherent field and if it is  strongly damped at the rate
$\kappa$, the field state will relax to  approximately a coherent state with
amplitude $\alpha=\frac{-2iE_{0}}{\kappa}$.   

We assume that $E_{0}/\kappa  \ll 1$. Then we can transform the total state by 
\begin{equation}
\tilde{\Lambda}=D^{+}(\alpha)\Lambda D(\alpha). 
\end{equation} 
Therefore $a\rightarrow a+\alpha$ and $a^{+}\rightarrow a^{+}+\alpha^{*}$.  
We then expand $\tilde{\Lambda}$\cite{Wiseman} 
\begin{equation}
{\tilde{\Lambda}}=\rho_{0}\otimes|0\rangle_{a}\langle 0|
+(\rho_{1}\otimes|1\rangle_{a}\langle 0|+H.c.)
+\rho_{2}\otimes|1\rangle_{a}\langle 1|
+({\rho^{'}}_{2}\otimes|2\rangle_{a}\langle 0|+H.c.). 
\end{equation}   

The reduced density operator is $\rho=Tr(\tilde{\Lambda})=\rho_{0}+\rho_{2}$ 
and the master equation after adiabatic elimination is 
\begin{equation}
\frac{d\rho}{dt}=\frac{1}{i\hbar}[\hat{H}_{0},\rho]-D[\hat{J},[\hat{J},\rho]].
\end{equation} 
Here 
\begin{equation} 
D=\frac{2g^4E_{0}^2}{\Delta^2\kappa^3}
\end{equation} 
is the diffusion constant and 
\begin{equation}
\hat{J}=-\cos(2k_{L}\hat{x}), 
\end{equation} 

\begin{equation}
\hat{H}_{0}=\frac{\hat{p}^2}{2M}+\hbar \chi \hat{J}, 
\end{equation} 
where
\begin{equation} 
\chi=\frac{2g^2E_{0}^2}{\Delta\kappa^2}. 
\end{equation}   

The conditional master equation for the optical field undergoing continuous
Homodyne measurement is\cite{Wiseman,Corney} 

\begin{equation}
(\frac{d\rho_{c}}{dt}_{field})=
\frac{\kappa}{2}(2a\rho_{c}a^{+}-a^{+}a\rho_{c}-\rho_{c}a^{+}a)+
\sqrt{\kappa}\frac{dW(t)}{dt}(a\rho_{c}+\rho_{c}a^{+}- \langle
a+a^{+}\rangle_{c}\rho_{c}), 
\end{equation} 

where $dW(t)$ is the infinitesimal Wiener increment. In this equation 
$\rho_{c}$ is the density matrix that is
conditioned on a particular realization of the Homodyne current up to time
$t$.   
The corresponding stochastic Schr\"{o}dinger equation is 
\begin{equation}
d|\psi_{c}(t)\rangle=dt[-i\hat{H}_{eff}/\hbar-\frac{1}{2}\kappa a^{+}a +
I(t)a]|\psi_{c}(t)\rangle, 
\end{equation}  
where 
\begin{equation}
I(t)=\kappa\langle a+a^{+}\rangle+\sqrt{\kappa}\frac{dW(t)}{dt}  
\end{equation}
is the measured current. Using Eq. (7), we can derive the nonlinear stochastic
Schr\"{o}dinger  equation by adiabatic elimination, 
\begin{equation}
d|\psi_{c}(t)\rangle=dt[-{i\hat{H}_{0}}/{\hbar}-D\hat{J}^2+
I_{A}(t)\hat{J}]|\psi_{c}(t)\rangle, 
\end{equation}   
where $I_{A}=4D\langle\hat{J}\rangle+\sqrt{2D}\frac{dW(t)}{dt}$.   

The normalized nonlinear
stochastic Schr\"{o}dinger equation is  
\begin{equation}
d|\psi_{c}(t)\rangle=dt[-{i\hat{H}_{0}}/{\hbar}-D(\hat{J}-\langle\hat{ J}
\rangle_c)^2+ \sqrt{2D}(\hat{J}-\langle
\hat{J}\rangle_c)\frac{dW(t)}{dt}]|\psi_{c}(t)\rangle. 
\end{equation}    

Given the modulation frequency $\omega$, we can define dimensionless  
parameters by $\tilde t=\omega t$,  $\tilde p=(\frac{2k_{L}}{M\omega})p$, 
$\tilde x=2k_{L}x$, $\tilde{\hat {H}}_0=\frac{4k^2_L}{M\omega^2}\hat {H}_0$, 
$\tilde g=g/\omega$, $\tilde E=E/\omega$, $\tilde \Delta=\Delta/\omega$,   
$\tilde \kappa=\kappa/\omega$,  $\tilde D=D/\omega$, and $\tilde  
\chi=\chi/\omega$.
This yields the commutator relation  
\begin{equation} 
[\tilde {\hat{x}}, \tilde
{\hat{p}}]=i\kbar, 
\end{equation} 
where $\kbar=\frac{4\hbar k^2_L}{M\omega}$ is
the dimensionless Plank constant.  

Omitting all the tildes , the equivalent 
equations are similar  except $\hbar$ is replaced by $\kbar$ and the 
dimensionless Hamiltonian is 
\begin{equation}
\hat{H}_{0}=\frac{\hat{p}^2}{2}-\xi \cos{\hat{x}}, 
\end{equation} 
where
$\xi=\frac{4k^2_L}{M\omega^2} \hbar\chi$.  In order to study  chaos in a
quantum system, we consider a periodic  modulation of the driving field
$E_{0}(t)=E_0\sqrt{1-2\epsilon\cos{t}}$.  The expressions of the stochastic
equations (14) and (15) will not change except that $\hbar$  is replaced by 
$\kbar$ and  $D$ replaced by $D(1-2\epsilon\cos{t})$ and $\xi$ in Eq. (17) is
replaced by $\xi (1-2\epsilon\cos{t})$.  

\section{sensitivity to diffusion constant} 

We assume that initially the atomic center-of-mass wave function is
in a Gaussian minimum uncertainty state with the position representation
\begin{equation}
\psi(x)=\left(\frac{1}{2\pi\sigma_x}\right)^{1/4}\exp[{-\frac{(x-x_{0})^2}
{4\sigma_x}+\frac{ip_{0}x}{\kbar}}]. 
\end{equation} 
We take $x_0=0$, $p_0=1.0$ as for these
values the state is localized on a second order period one resonance and is
thus localized in a regular region of phase space (see  Fig. 1).  For
$\sigma_x=0.3906$, $\kbar=0.25$, $\xi=1.2$,  Dyrting et al. \cite{Dyrting} 
have
shown that the system will coherently tunnel between the two corresponding
second order period one resonances.  We use a Split Operator Method
\cite{Kosloff} and  FFT (Fast Fourier Transformation) \cite{Press} to obtain
the numerical solution of the stochastic Schr\"{o}dinger equations. In this
scheme the kinetic operator and potential operator are used separately to
propagate the wave function:  
\begin {equation} \exp[-i\hat{\bf
H}\/{\delta}t/\kbar]\sim \exp[-i(\hat{\bf P}\/)^2{\delta}t/{4\kbar}]
\exp[-i(\hat{\bf V}\/){\delta}t/{2\kbar}] \exp[-i(\hat{\bf
P}\/)^2{\delta}t/{4\kbar}]. 
\end{equation} 
The computing errors are of
$O(\delta t^3)$. Here $\hat{V}$ is the effective potential which includes a
stochastic term, 
\begin {equation} \hat{V}=-\xi(t)\cos{\hat{x}}-i\kbar
[D(t)(\hat{J}-\langle\hat{ J} \rangle_c)^2(1+\frac{dW(t)^2}{dt})-
\sqrt{2D(t)}(\hat{J}-\langle \hat{J}\rangle_c)\frac{dW(t)}{dt}].
\end{equation}  
where $\xi(t)=\xi(1-2\epsilon\cos{t})$ and
$D(t)=D(1-2\epsilon\cos{t})$. We introduced the $\frac{(dW(t))^2}{dt}$  term to
keep the expression consistent with the normalized nonlinear  stochastic
Schr\"{o}dinger equation  after an expansion of the exponential function.  

In
order to compare the quantum and semi-classical stochastic evolutions, we  
calculate  the Wigner function\cite{Wigner},\cite{Wang}   
\begin{equation}
P(x,p)=\frac{1}{2\pi\kbar}\int{dy\langle{x-\frac{y}{2}|\rho|x+\frac{y}{2}}}
\rangle \exp({ipy/\kbar}). 
\end{equation} 
This expression can be interpreted as
the Weyl-Wigner correspondence \cite{Wang}  of the density operator. To give
the  dynamical equation for the Wigner function that is the quantum
correspondence  of a classical Liouville equation we use  the Weyl-Wigner
correspondence of an operator $\hat{F}=\hat{A}\hat{B}$  which is
\cite{Groenewold,Wang} 
\begin{equation} 
F(x,p)=A(x,p)XB(x,p), 
\end{equation}
where $X=\exp[\frac{\kbar}{2i}(\frac{\lvec}{\partial p}\frac{\rvec}{\partial
x}- \frac{\lvec}{\partial x}\frac{\rvec}{\partial p})]$ and the arrows on the
operators denote the  term on which the operator is to be applied.
Alternatively, we obtain 
\begin{equation}
F(x,p)=A(x-\frac{\kbar}{2i}\frac{\partial}{\partial p},
p+\frac{\kbar}{2i}\frac{\partial}{\partial q})B(x,p). 
\end{equation}  

When we
apply this formula to the products appearing in the Master equation  we can
readily obtain the phase space equation for the Wigner function we are looking
for 
\begin{equation} \frac{\partial P}{\partial t}=[\frac{\partial
H_{0}(t)}{\partial q}\frac{\partial P} {\partial p}-\frac{\partial
H_{0}(t)}{\partial p}\frac{\partial P}{\partial q}]
+D(t)\kbar^2\sin^2{x}\frac{\partial^2 P}{\partial p^2}, 
\end{equation} 
where
$H_0(t)$ is the classical Hamiltonian including modulation
${H}_{0}(t)=\frac{{p}^2}{2}-\xi(t) \cos{{x}}$.  We give the corresponding
classical stochastic F-P equation from quantum nonlinear stochastic 
Schr\"{o}dinger equation\cite{Walls} 
\begin{equation} 
\frac{dx}{dt}=p,
\end{equation}  

\begin{equation}
\frac{dp}{dt}=-\xi(t)\sin{x}+\sqrt{2D(t)}\kbar\sin{x}\frac{dW(t)}{dt}.
\end{equation}   
To describe the classical distribution we use the classical
$Q$ function   \cite{Chen}. The initial state is a bivariate Gaussian centered
on $(x_0, p_0)$  with position variance $\delta_x$ and momentum variance
$\delta_p$,   

\begin{equation}
Q_{0}(x,p)=\frac{1}{2\pi\sqrt{\delta_{x}\delta_{p}}}
\exp[-\frac{(p-p_{0})^2}{2\delta_p}] \exp[-\frac{(x-x_{0})^2}{2\delta_x}],
\end{equation}  
where the classical variances $\delta_x$ and $\delta_{p}$ are
related with quantum parameters
$\delta_{x}=\frac{\kbar^2}{2\xi}+\frac{\kbar^2}{4\sigma_x}$,
$\delta_{p}=\frac{\kbar\sqrt{\xi}}{2}+\sigma_p$. 
The evolution of $Q$ function
is $Q(x, p, t)=Q_{0}[\bar{x}(x, p, -t), \bar{p}(x, p, -t)]$, where $\bar{x}(x,
p, -t), \bar{p}(x, p, -t)$ is the trajectory generated by Hamilton's
equations.  

To compare the quantum dynamics with the classical conditional
dynamics,  for quantum system we study the  ensemble  with the same initial
condition but with random trajectories. It shows that when $D$ is very small,
for Homodyne measurement, the  evolution of  average momentum
$\langle{p}\rangle$ and  average variance of momentum 
$(\langle{p^2}\rangle-\langle{p}\rangle^2)$ for the ensemble  show  coherent
tunneling. Therefore the perturbation is not serious for small $D$ when the
initial state is in the regular region of the classical phase space.  

 For the
classical dynamics for small $D$ the results are close to the  no diffusion
case\cite{Dyrting}.   Obviously we therefore  expect that we obtain different
results  between   classical and quantum conditional dynamics (Fig. 2). 

However, if the diffusion constant is large enough, we obtain almost the  same
result as in the classical case (Fig. 3).  For a single stochastic
measurement,  the terms in the normalized nonlinear Schr\"{o}dinger 
stochastic  equation  due to the measurement depend on the quantity 
$(\hat{J}-\langle \hat{J}\rangle_c)$. We expect that for some range of  values
of $D$, the stochastic measurement terms would drive the system towards  an
oscillating trajectory for which  

\begin{equation}  
\langle \hat{J}^2\rangle_c
\simeq \langle \hat{J}\rangle^{2}_c.  
\end{equation} 
Therefore for the ensemble
which includes many random trajectories, the results will approach  that of the
classical conditional dynamics.  

\section{Sensitivity to chaotic and regular initial states}   

We again assume that the wave  function is initially in a
minimum uncertainty state in the position representation. We choose two initial
locations in classical phase space. In the first case, $x_0=0$, $p_0=1.0$  it
is in the regular region of classical phase space. In the second case
$x_0=-2.5$, $p_0=1.0$, it is in the chaotic region (see Fig. (1)).  

Because the
measurement will perturb the quantum state, we hope to be able to compare the
effect of  measurement noise of different trajectories. Here the angle
$\theta_{ij}(t)$ is defined between  two normalized state vectors
$|\psi_{i}(t)\rangle$ and $|\psi_{j}(t)\rangle$ as\cite{Schack,Breslin}
\begin{equation}
\theta_{ij}(t)=\cos^{-1}|\langle\psi_{i}(t)|\psi_{j}(t)\rangle|.
\end{equation}  
In the position representation, this is 
\begin{equation}
|\langle\psi_{i}(t)|\psi_{j}(t)\rangle|=|\int_{-\infty}^{\infty}
\psi_{i}(x,t)^*\psi_{j}(x,t)dx|. 
\end{equation} 
As a measure of the
distribution of the state vectors in Hilbert space we can calculate the average
angle between all pairs of vectors. We define 
\begin{equation}
\theta_{ave}(t)=\frac{2}{(N^2-N)}{\sum_{i \neq j} {\theta_{ij}(t)}},
\end{equation} where $N$ is the number of trajectories.     

In Fig. 4, we plot
${\theta}_{ave}(t)$ for the two initial states mentioned  above, evolved up to
200 cycles.   We used up to 40000 steps and $N=1000$ trajectories for our 
calculation. As can be seen, the average angle between vectors 
starting in the chaotic phase space region is larger than that of the  regular
initial state.   

If the system started in the regular region, small errors in
recording the results of the measurement will not be a very serious problem
because the conditional states form trajectories which will remain close in
Hilbert space. Therefore, if we consider the distribution of Hilbert angles at
a fixed strobe  number(Fig. 5),  the distribution is centered at a
small angle for an initial regular state. On the other hand, for an initial
chaotic state, the peak location approaches $\pi/2$ (Fig. 6). It means that 
for an initial chaotic state most vectors are far apart  from each other.
Therefore if the initial state was in the chaotic region of phase space, it
will be much more difficult to infer the system state reliably from
the measurement results. This result is consistent with the results of the
quantum kicked top\cite{Breslin}.   

In summary, we have demonstrated that for
continuous Homodyne measurement of signals from the quantum system of single
atom dynamics in cavity QED, the measured results are  influenced by the
diffusion constant and  whether the initial states are in regular or chaotic
phase space  regions.  We have shown that if the diffusion constant is large
enough, the average measurement results are similar to classical
conditional dynamics and  for small diffusion constant the initial chaotic
state will  be more sensitive to errors in recording the measurement results
than the initial regular state.  

\section{Acknowledgment} 

One of authors(XML)
would like to thank Dr. J. F. Corney and Dr. M. Gagen  for useful discussions
on  conditional master equations and Homodyne measurements.

\newpage 
%FFFFFFFFFFFFFFFFFFFFFFFFFFFFFFFFFFFFFFFFFFF 
\begin{figure}[htbp]
\caption{{\em  Stroboscopic portrait of the system with Hamiltonian
$H_{0}=\frac{p^2}{2}-\xi(1-2\epsilon\cos{t})\cos{x}$, where $\epsilon=0.2$,
$\xi=1.2$.} } \protect\label{fig_NM} 
\end{figure}
%FFFFFFFFFFFFFFFFFFFFFFFFFFFFFFFFFFFFFFFFFFF  

%FFFFFFFFFFFFFFFFFFFFFFFFFFFFFFFFFFFFFFFFFFF 
\begin{figure}[htbp]
\caption{{\em  Evolution of  average momentum $\langle{p}\rangle$ and and
average variance of momentum  $\langle{p^2}\rangle-\langle{p}\rangle^2$ for
classical and quantum  conditional dynamics when $D=0.001$. 1000 random
trajectories are taken. Solid  line, classical conditional dynamics, dashed
line, quantum conditional  dynamics.} } \protect\label{fig_NM} 
\end{figure}
%FFFFFFFFFFFFFFFFFFFFFFFFFFFFFFFFFFFFFFFFFFF 

%FFFFFFFFFFFFFFFFFFFFFFFFFFFFFFFFFFFFFFFFFFF 
\begin{figure}[htbp]
\caption{{\em  Evolution of  average momentum $\langle{p}\rangle$ and and
average variance of momentum  $\langle{p^2}\rangle-\langle{p}\rangle^2$ for
classical and quantum  conditional dynamics and $D=0.1$. 1000 random
trajectories are taken. Solid  line, classical conditional dynamics, dashed
line, quantum conditional  dynamics.} } \protect\label{fig_NM} 
\end{figure}
%FFFFFFFFFFFFFFFFFFFFFFFFFFFFFFFFFFFFFFFFFFF 

%FFFFFFFFFFFFFFFFFFFFFFFFFFFFFFFFFFFFFFFFFFF 
\begin{figure}[htbp]
\caption{{\em  Evolution of  average angles in Hilbert space for $D=0.001$.
Solid line, in chaotic region initially $x_{0}=-2.5$, $p_{0}=1.0$. Dashed line,
in regular region initially $x_{0}=0.0$, $p_{0}=1.0$. } }
\protect\label{fig_NM} 
\end{figure}
%FFFFFFFFFFFFFFFFFFFFFFFFFFFFFFFFFFFFFFFFFFF 

%FFFFFFFFFFFFFFFFFFFFFFFFFFFFFFFFFFFFFFFFFFF 
\begin{figure}[htbp]
\caption{{\em  Distribution of angles  in Hilbert space at strobe number 200
for initial regular state and $D=0.001$. } } \protect\label{fig_NM}
\end{figure} 
%FFFFFFFFFFFFFFFFFFFFFFFFFFFFFFFFFFFFFFFFFFF 

%FFFFFFFFFFFFFFFFFFFFFFFFFFFFFFFFFFFFFFFFFFF 
\begin{figure}[htbp]
\caption{{\em  Distribution of angles  in Hilbert space at strobe number 200
for initial chaotic  state and $D=0.001$. } } \protect\label{fig_NM}
\end{figure} 
%FFFFFFFFFFFFFFFFFFFFFFFFFFFFFFFFFFFFFFFFFFF  


\thebibliography{99}  

\bibitem{Zurek94}  W.H.Zurek and J.P.Paz, Phys. Rev.
Lett. {\bf 72}, 2508 (1994). 
 
\bibitem{Schack94}R.Schack, G.M.D'ariano and
C.M.Caves, Phys. Rev E {\bf 50}, 972 (1994).    
\bibitem{Fishman82}S.Fishman,
D.R.Gremple and R.E. Prange, Phys. Rev. Lett. {\bf 49}, 509 (1982). 
\bibitem{Shepelyansky86}D.L.Shepelyansky, Phys. Rev. Lett. {\bf 56}, 677
(1986).  

\bibitem{Chirikov89} B.V.Chirikov, F.M.Israilev, D.L.Shepelyansky,
Physica D {\bf 33}, 77 (1989).  

\bibitem{Zurek} W.H.Zurek, Physica Scripta {\bf
T76},  186, (1998).  

\bibitem{Galvez88} E.J.Galvez, B.E.Sauer, L.Moorman,
P.M.Koch and D. Richards, Phys. Rev. Lett. {\bf 61}, 2011 (1988). 

\bibitem{Bayfield89} J.E.Bayfield, G.Casati, I.Guarneri and D.W.Sokol, Phys.
Rev. Lett. {\bf 63}, 364 (1989).  

\bibitem{Arndt91} M.Arndt, A. Buchleitner,
R.N. Mantegna and H.Walther, Phys. Rev. Lett. {\bf 67}, 2435 (1991). 

\bibitem{Moore95}  F.L. Moore, J.C. Robinson, C.F. Bharucha, Bala Sundaram, and
M.G. Raizen, Physical Rev. Lett. {\bf 75}, 4598, (1995).  

\bibitem{Klappauf1}
B. G. Klappauf, W. H. Oskay, D. A. Steck, and M. G. Raizen, Physical Rev. Lett.
{\bf 81}, 1203, (1995).  
 
\bibitem{Klappauf2} B. G. Klappauf, W. H. Oskay, D.
A. Steck, and M. G. Raizen, Physical Rev. Lett. {\bf 81}, 4044, (1995). 

\bibitem{Ammann98} H.~Ammann, R.Gray, I~Shvarchuck, and N.~Christensen.
Physical Rev. Lett. {\bf 80}, 4111 (1998). 
  
\bibitem{Ye} J.Ye, D.W.Vernooy,
and H.J.Kimble, Phys. Rev. Lett. {\bf 83}, 4987(1999).  

\bibitem{doherty99}
A.C.Doherty, S.M.Tan, A.S.Parkins and D.F.Walls, Phys. Rev A {\bf 60}, 2380
(1999).   

\bibitem{Schack} R\"{u}diger Schack, and Carlton M. Caves, Phys. Rev.
A {\bf 53}, 3257(1996).  

\bibitem{Breslin} J.K. Berslin, and G. J. Milburn,
Phys. Rev. A {\bf 59}, 1781(1999).  

\bibitem{carmichael}H.J.Carmichael, {\em An
Open Systems Approach to Quantum Optics}, Lecture Notes in Physics: New Series
m, Monographs, {\bf m18}, (Springer, Berlin, 1993).  

\bibitem{molmer} K.
Molmer, Y.Castin, Quantum Semiclass. Opt. {\bf 8}, 49 (1996).  

\bibitem{Brun97}
T.A.Brun, Gisin N, P.F.O'Mahony , M. Rigo, Phys. Lett A {\bf 229}, 267(1997).
 
\bibitem{Herkommer} A.M.Herkommer, H.J.Carmichael and W.P.Schleich,   Quantum
Semiclass. Opt. {\bf8}, 189(1996).  

\bibitem{Quadt}  Ralf Quadt, Matthew
Collett, and Dan F. Walls,   Phys. Rev. Lett. {\bf 74}, 351(1999). 

\bibitem{mabuchi} H. Mabuchi, J.Ye, H.J.Kimble, Applied Physics B-Lasers and
Optics {\bf 68},1095,(1999).  

\bibitem{Wiseman} H. M. Wiseman, and G. J.
Milburn, Phys. Rev. A {\bf 47}, 642(1993).  

\bibitem{Corney} J. F. Corney, and
G. J. Milburn, Phys. Rev. A {\bf 58}, 2399(1998).  

\bibitem{Dyrting} S.
Dyrting, and G. J. Milburn, Phys. Rev. A {\bf 51}, 3136(1995). 

\bibitem{Kosloff} Ronnie Kosloff, J. Phys. Chem. {\bf 92}, 2087(1988). 

\bibitem{Press} William H Press et al (1992), \em Numerical Recipes in C \em,
Cambridge, University Press.  

\bibitem{Wigner} E. Wigner, Phys. Rev. {\bf 40},
749(1932). 
 
\bibitem{Groenewold} H. J. Groenewold, Physica {\bf 12},
405(1946).  

\bibitem{Wang} L. Wang and R. F. O'Connell, Found. Phys. {\bf 18},
1023(1988).  

\bibitem{Walls} D. F. Walls, G. J. Milburn(1994), \em Quantum
Optics\em, Springer-Verlag Berlin Heidelberg. 

\bibitem{Chen} Wenyu Chen, S.
Dyrting, and G. J. Milburn, Aust. J. Phys. {\bf 49}, 777(1996).
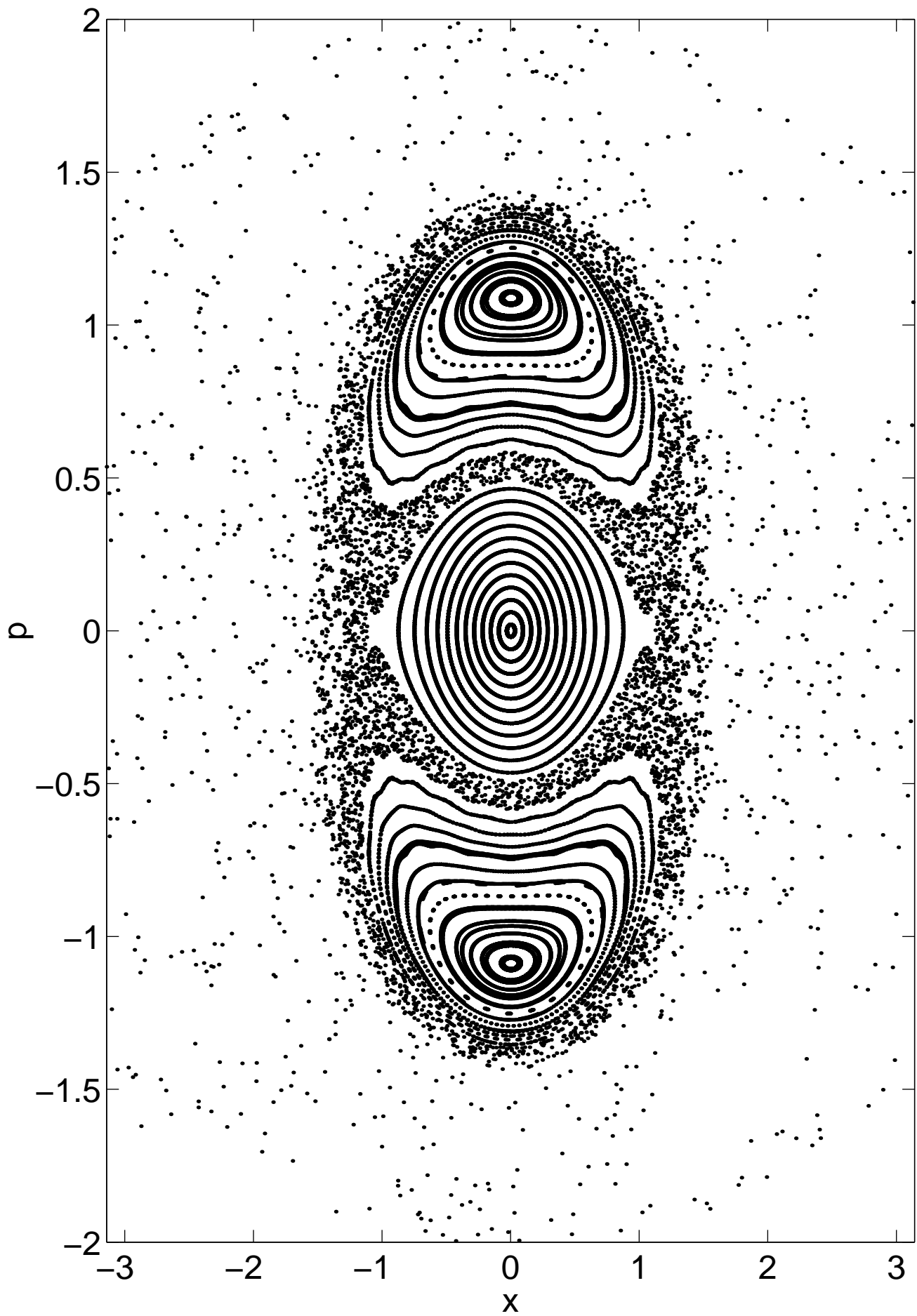
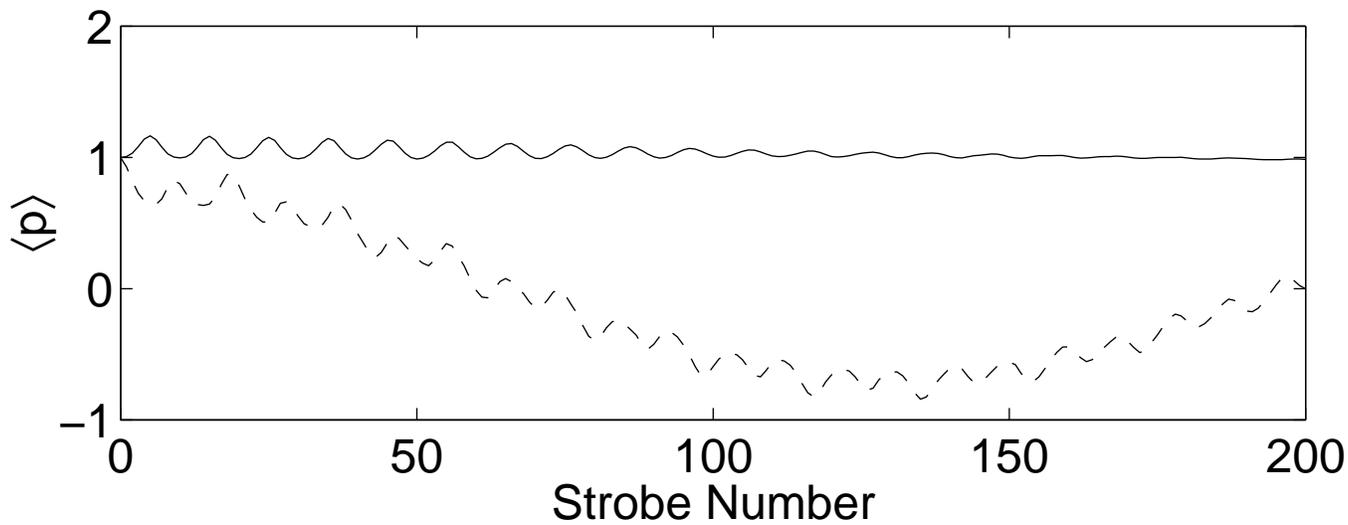
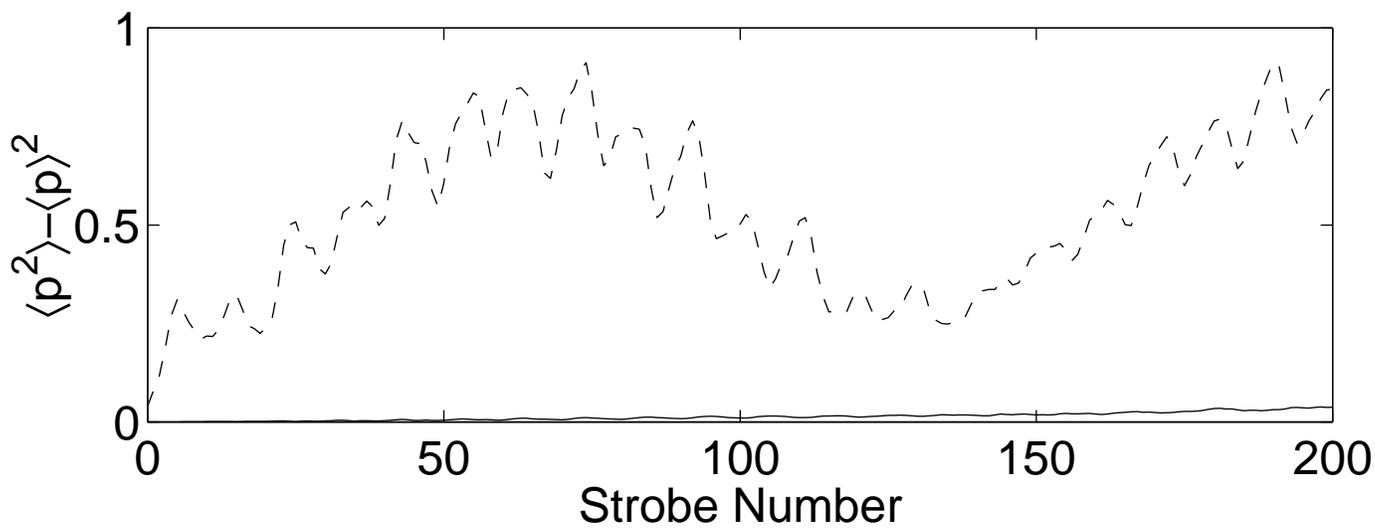
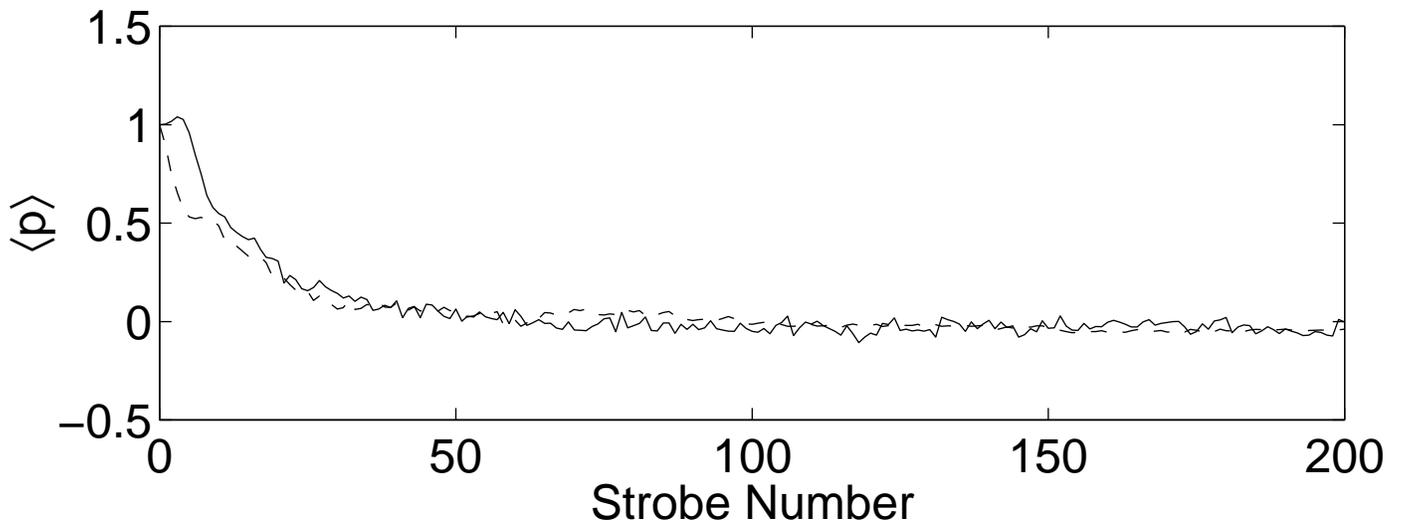
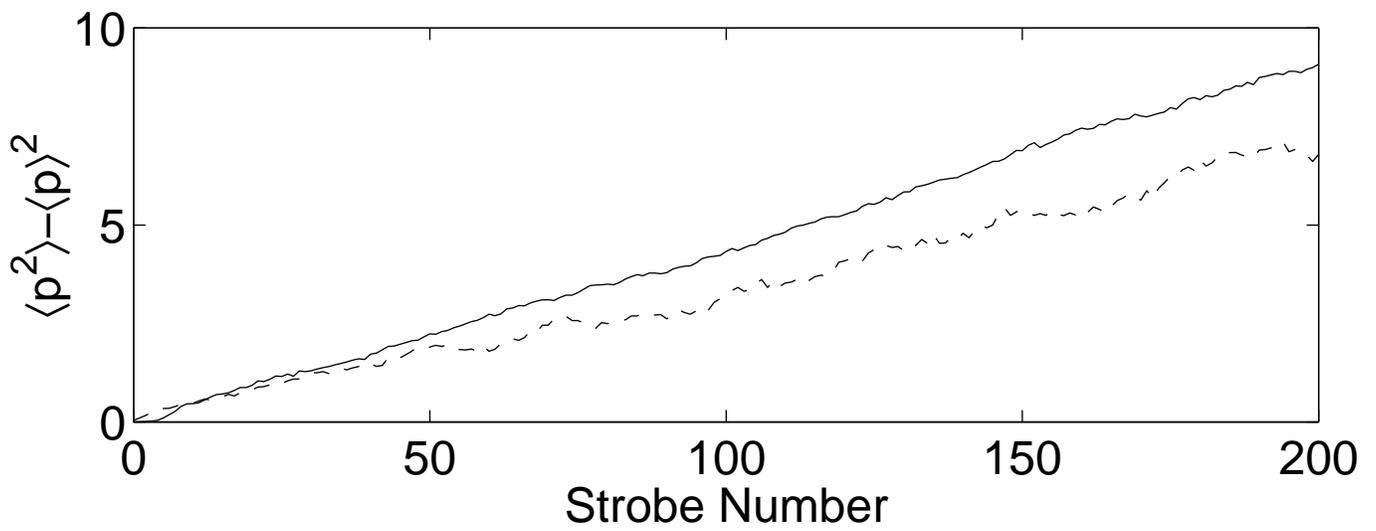
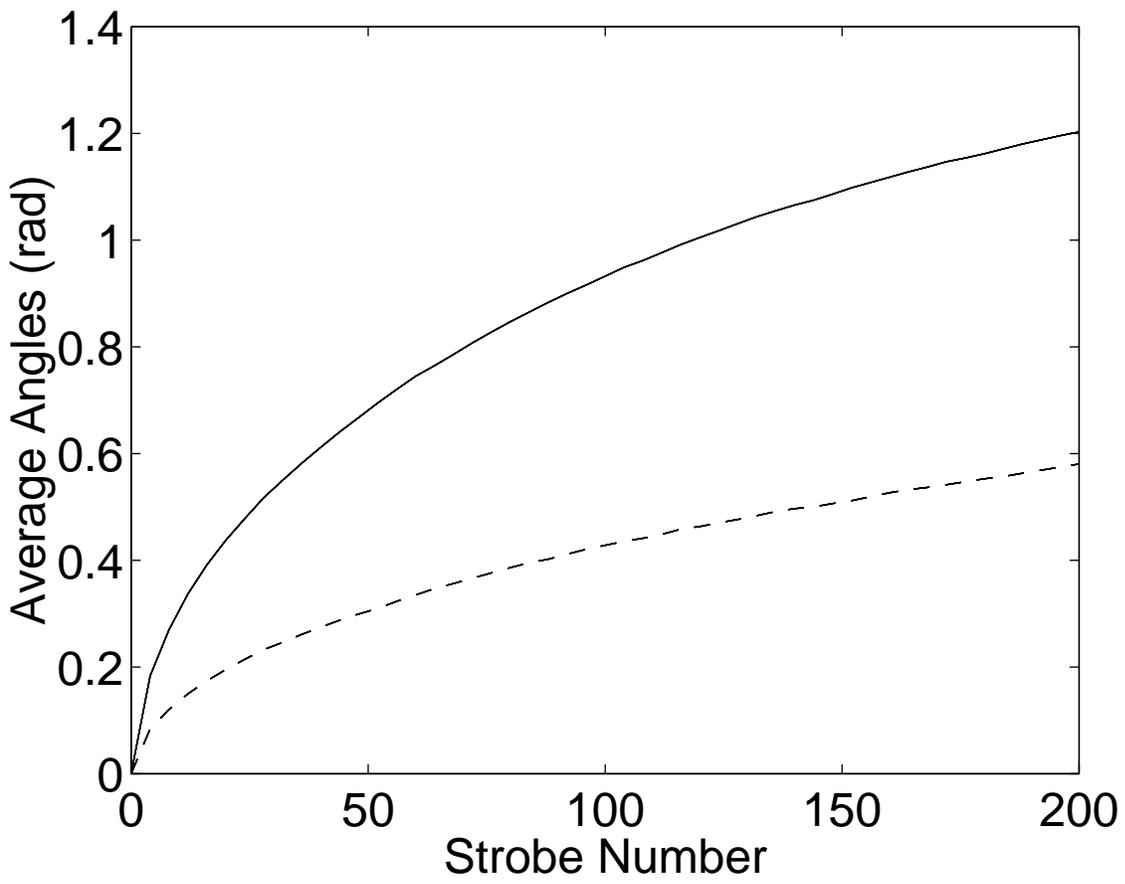
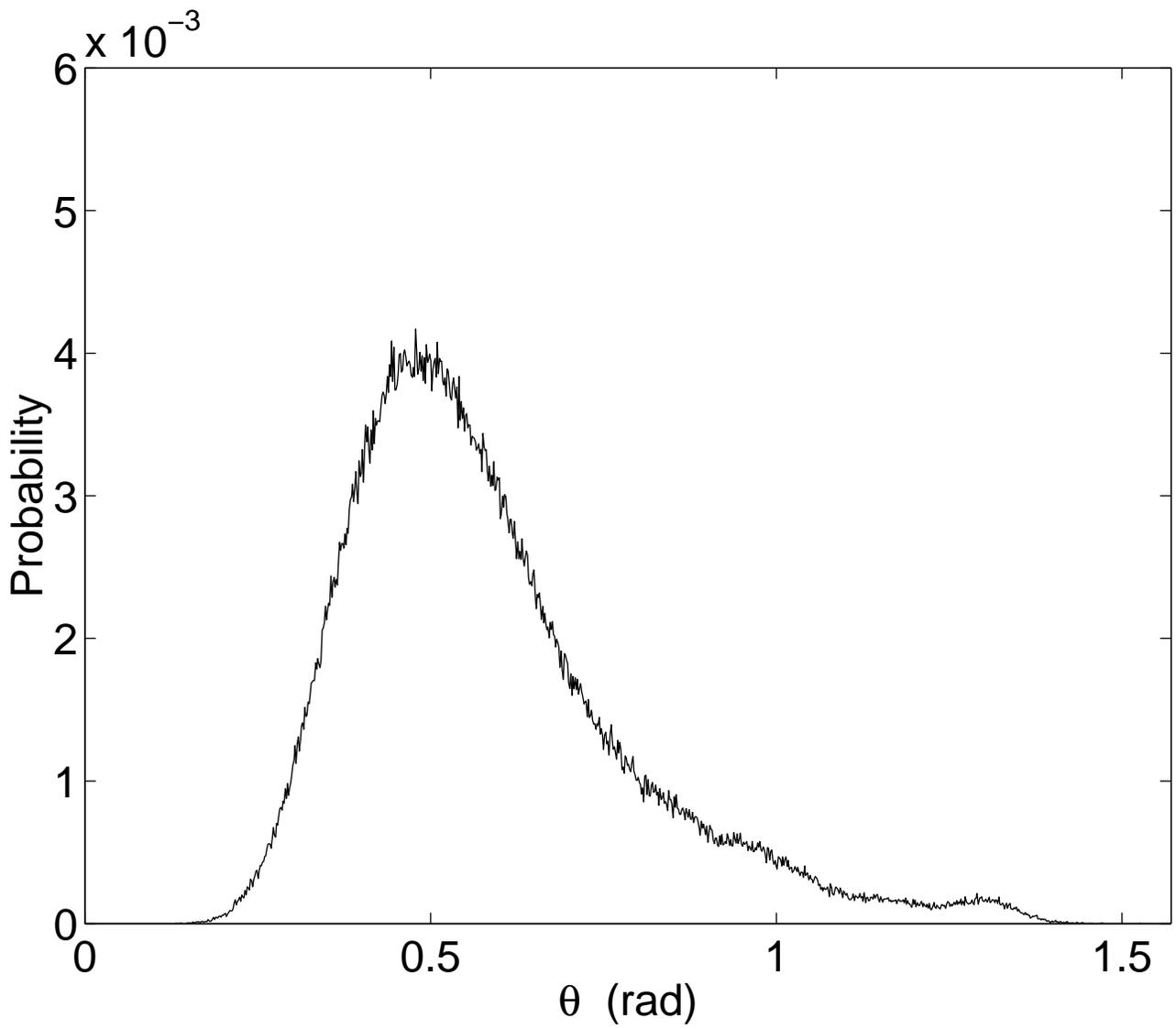
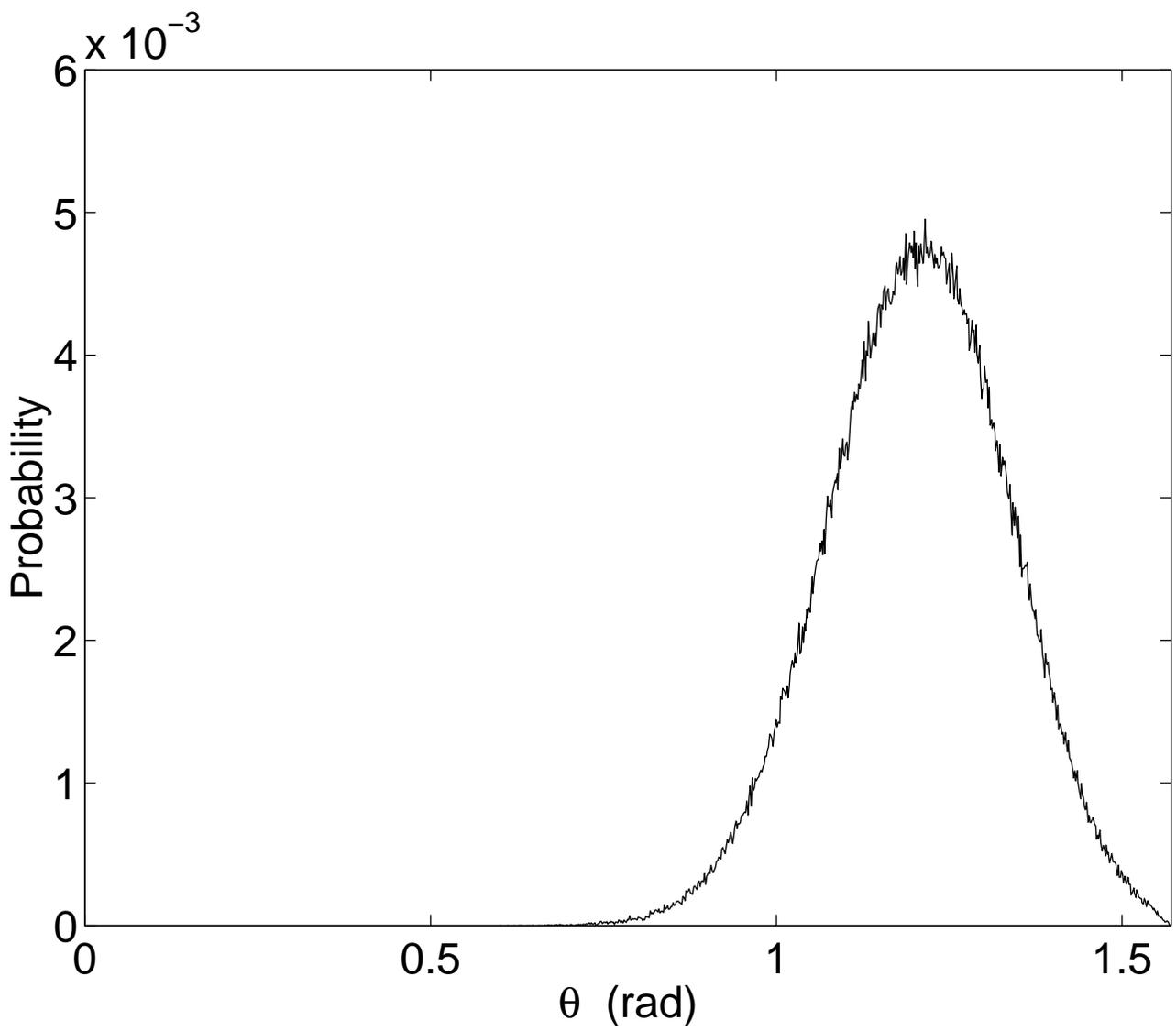
\end{document}